\documentclass[a4paper,11pt]{article}
\pdfoutput=1 

\usepackage{jheppub} 
\usepackage{xspace}
\usepackage{booktabs}
\usepackage[utf8]{inputenc}                    
\usepackage[T1]{fontenc} 
\usepackage{amsmath}

\newcommand\optional[1]{}
\usepackage{color}

\newcommand{\seff}{\ensuremath{\sigma_{\!\textit{eff}}}\xspace}

\title{\boldmath Extracting \seff from the LHCb double-charm measurement}

\author[a]{M. H. Seymour}
\author[a]{A. Si\'odmok}

\affiliation[a]{School of Physics and Astronomy, 
		\\ The University of Manchester, Manchester, M13 9PL, U.K.}

\emailAdd{michael.seymour@manchester.ac.uk}\emailAdd{andrzej.siodmok@manchester.ac.uk}

\abstract{We discuss various issues related to the definition of single
  and double open charm cross sections. We conclude that LHCb's
  extraction of \seff, the effective cross section for double-parton
  scattering, is too large by a factor of two.
  This correction brings the data from open-charm pairs closer to that
  from $\mathrm{J}/\psi$ plus open charm and jet production.
}
\preprint{{\flushright MAN/HEP/2013/20\\ MCnet-13-11\\ }}
\begin{document}

\maketitle
\flushbottom
\section{Introduction}
Multi-parton interactions are firmly established as the primary source
of underlying event activity in high energy hadronic collisions (see for
example Ref.~\cite{Abramowicz:2013iva} and references therein). However,
attempts to study their properties by cleanly identifying multi-
(initially double-) parton scattering events have proved
difficult.

The general principle behind all such measurements is to assume that
there is little correlation between the two scatters, so that using
final state observables one can separate the signal for a given process
into a correlated contribution coming from one scatter and two
uncorrelated components each coming from a different scatter. The
earliest searches for and measurements of double-parton scattering used
four jet production (AFS~\cite{Akesson:1986iv}, UA2~\cite{Alitti:1991rd}
and CDF~\cite{Abe:1993rv}), in which the double-parton signal consists
of two back-to-back dijet pairs, uncorrelated in azimuth, whereas the
single-parton background consists of a back-to-back dijet pair, with two
additional jets produced by initial- or final-state bremsstrahlung; the
former giving little correlation between the two emissions, the latter
giving a strong correlation with the primary jet pair.

The most precise measurements of double-parton scattering to date come
from the Tevatron measurements of $\gamma+3$jets production
(CDF~\cite{Abe:1997xk} and D0~\cite{Abazov:2009gc}), based on the same
idea, but with one of the jets replaced by a photon, which has a
considerably smaller single-parton background.

The LHC measurements have focussed on channels that have even smaller
single-parton backgrounds, for example W~$+$~dijets
(ATLAS~\cite{Aad:2013bjm}, with preliminary work towards such a
measurement also by CMS~\cite{CMS:awa}), in which again the dijet pair
should be back-to-back, with little recoil from the W. The ultimate
channel in this direction would be like-sign W pair
production\cite{Kulesza:1999zh}, for which not only is the single-parton
background small, but it is also very distinctive since by charge
conservation the W pair must be accompanied by at least two high-$p_t$
jets.

LHCb have studied double-parton scattering by measuring double-charm
cross sections. In particular, they have measured double-charmonium
production in~\cite{Aaij:2011yc}, charmonium~$+$~open-charm production
in~\cite{Aaij:2012dz} and double-open-charm production, also
in~\cite{Aaij:2012dz}. The channels involving charmonia, as well as
double-open-charm production in the case that both of the measured
charmed hadrons contain a charm quark (or both an anticharm), are
expected to be dominated by double-charm production. On the other hand,
double-open-charm production in which one charmed hadron contains a
charm quark and the other an anticharm is expected to be dominated by
single-charm production, although with a significant contribution from
double-charm production that could perhaps be separated using
correlations in their phase space distributions.

The amount of double-parton scattering is typically parametrized by the
effective cross section, \seff, through:
\begin{equation}
  \label{eq:sigmaeffdef}
  \sigma_{\mathrm{ii}} = \frac{\sigma_{\mathrm{i}}^2}{2\seff}\,, \qquad
  \sigma_{\mathrm{ij}} = \frac{\sigma_{\mathrm{i}}\sigma_{\mathrm{j}}}{\seff}\,,
\end{equation}
where $\sigma_{\mathrm{i}}$ and $\sigma_{\mathrm{j}}$, and
$\sigma_{\mathrm{ii}}$ and $\sigma_{\mathrm{ij}}$, are suitably defined
cross sections for single- or double-parton scatters of types
$\mathrm{i}$ and $\mathrm{j}$. We
discuss their definitions in more detail in the next section, but when
they are defined properly, we take Eq.~(\ref{eq:sigmaeffdef}) as
defining \seff. The factor of~2 appearing in Eq.~(\ref{eq:sigmaeffdef})
is a simple symmetry factor.

In this paper, we discuss LHCb's extraction of \seff from their
measurements of single- and double- open-charm production cross
sections. We wish to stress that we do not question their measurements
of the cross sections themselves, only the way they combine them to
extract \seff, which has the potential to act as a strong constraint 
on models of multiple parton scattering and, in particular, 
their models of the transverse-space distribution of
partons in hadrons \cite{Seymour:2013qka}.

The remainder of the paper is as follows. In
Section~\ref{sect:inclusiveCrossSections} we discuss the cross
section definitions for single- and double- inclusive cross sections.
We show that, in the eikonal model of multi-parton
interactions, only if the cross sections are defined in the correct,
inclusive, sense, is the effective cross section defined by
Eq.~(\ref{eq:sigmaeffdef}) a process-independent quantity. Technical
details are deferred to
Appendix~\ref{sect:inclusiveCrossSectionsDetails}.
In Section~\ref{sect:charmCrossSections}, we specialize these to the
cases of charm quark and charmed hadron production. 
This section
already contains our main result: LHCb's extraction of \seff from their
single- and double-open-charm cross sections is too large by a factor
of~2. 
In Section~\ref{sect:results} we summarize our re-extraction of
\seff from the LHCb data.
\optional{In Section~\ref{sect:MonteCarlo} we make a
brief Monte Carlo study of double-charm production and highlight some
issues that may be of interest for future measurements of this
process.}
In
Section~\ref{sect:summary} we briefly summarize our paper. Finally, in
Appendix~\ref{sect:Berezhnoy}, we make comparisons with a theory paper
by two LHCb authors and others\cite{Berezhnoy:2012xq}, which appears to
agree with the LHCb results.

\section{Single- and double-inclusive cross sections}
\label{sect:inclusiveCrossSections}

Considerable confusion has arisen concerning cross section definitions
for use in studies of double-parton scattering. The classic measurement
of CDF\cite{Abe:1997xk}, for example, used an exclusive definition and
therefore the \seff they extracted was not the usual one, but a
process-dependent approximation to
it, which was pointed out and corrected in~\cite{Treleani:2007gi,Bahr:2013gkj}.

In this section we define precisely what we mean by the inclusive cross
sections and draw comparisons with results using another common
definition of the word ``inclusive'' and with exclusive cross
sections. In order to motivate our definition, we discuss an eikonal
model of multi-parton interactions, but we stress that our definition is
completely independent of that model. However, we will see that in that
model, our definition of inclusive cross sections leads to an effective
cross section that is a property only of the colliding hadrons and not
of the process by which it is measured.

We therefore beginning by defining this eikonal model. We assume that
parton distribution functions factorize into a longitudinal momentum
part and transverse space part, and further that multi-parton
distribution functions factorize into products of single-parton
distribution functions. In the details of exclusive final states, this
approximation must fail, but it is believed to be a good approximation
for the distribution of number of hard scatters, at least, and is the
basis of all current multi-parton interaction models that describe LHC
underlying event data
. In this approximation, the cross section for $n$
partonic scatters of a type $\mathrm{i}$ can be written as a convolution
over impact parameter, $b$, of a factor that represents the
Poisson-distributed probability of having $n$ independent collisions,
with $b$-dependent average value:
\begin{equation}
  \label{eq:eikonal}
  \sigma_{n\mathrm{i}} = \int \mathrm{d}^2b \,
  \frac{\left(\sigma_{\mathrm{i}}A(b)\right)^n}{n!} \,
  \mathrm{e}^{-\sigma_{\mathrm{i}}A(b)} \,,
\end{equation}
where $\sigma_{\mathrm{i}}$ is the cross section for a single-parton
scatter of type $\mathrm{i}$, calculated with the conventional
(inclusive) parton distribution functions, and $A(b)$ is referred to as
the matter distribution, normalized according to
\begin{equation}
  \int \mathrm{d}^2b \, A(b) = 1\,.
\end{equation}

\subsection{Single-particle cross sections}

\subsubsection{Inclusive definition}

The conventional definition of the inclusive production of some state
$\mathrm{i}$ is to imagine a hypothetical detector that counts
$\mathrm{i}$'s in some fiducial volume of phase space. The counter is
assumed to be perfect, in the sense that every $\mathrm{i}$ is counted,
independently of the structure of the event it appears in. In
particular, independently of whether the event contains additional
$\mathrm{i}$'s and, if so, how many. The number of $\mathrm{i}$'s
counted, $N_{\mathrm{i}}$, in a run with integrated luminosity
$\mathcal{L}$ then defines the inclusive $\mathrm{i}$ cross section:
\begin{equation}
  \label{eq:inclDefinition}
  \sigma_{\textit{incl\/}\mathrm{i}} \equiv
  \frac{N_{\mathrm{i}}}{\mathcal{L}}\,.
\end{equation}
The requirement that every $\mathrm{i}$ is counted, independently of how
many there are, is what is usually meant by inclusive. Note that this
definition does not require any reference to the number of events
measured, only the number of $\mathrm{i}$'s.

As shown in Appendix~\ref{sect:inclusiveCrossSectionsDetails}, this is
equivalent to writing
\begin{equation}
  \label{eq:inclLemma}
  \sigma_{\textit{incl\/}\mathrm{i}} = \sum_n n\,\sigma_{n\mathrm{i}}\,,
\end{equation}
where $\sigma_{n\mathrm{i}}$ is the exclusive cross section to produce
$n$ $\mathrm{i}$'s. Note that an event that contains $n$ $\mathrm{i}$'s
contributes $n$ times to the inclusive cross section. This may seem like
a simple result, but a surprising number of papers in both the theory
and experiment of multi-parton interactions do not make the step from
Eq.~(\ref{eq:inclDefinition}) to Eq.~(\ref{eq:inclLemma}).

The cross section for events in which there are exactly $n$
scatters of type $\mathrm{i}$ is precisely what we defined in the
eikonal model in Eq.~(\ref{eq:eikonal}). We can therefore write the
inclusive cross section in the eikonal model as
\begin{eqnarray}
  \sigma_{\textit{incl\/}\mathrm{i}} &=& \sum_n n\,\int \mathrm{d}^2b \,
  \frac{\left(\sigma_{\mathrm{i}}A(b)\right)^n}{n!} \,
  \mathrm{e}^{-\sigma_{\mathrm{i}}A(b)}
  = \int \mathrm{d}^2b \left[\sum_n n
  \frac{\left(\sigma_{\mathrm{i}}A(b)\right)^n}{n!}\right]
  \mathrm{e}^{-\sigma_{\mathrm{i}}A(b)}
  = \int \mathrm{d}^2b \, \sigma_{\mathrm{i}}A(b)
\nonumber\\
  &=& \sigma_{\mathrm{i}}\,.
\end{eqnarray}
That is, the inclusive cross section is equal to the partonic cross
section. This deceptively simple result is a non-trivial test of the
self-consistency of the eikonal model: since the partonic cross section
is calculated from the inclusive parton distribution functions, and
these are defined as operators for the production of a single parton
with all other information integrated out, it must be that this partonic
cross section is fully inclusive. That is, each parton that a hadron
produces is described by the parton distribution function and each
collision that that parton initiates contributes to the inclusive cross
section.

\subsubsection{Alternative inclusive definition}

Another definition of the word ``inclusive'' appears frequently in the
literature. In this definition, the inclusive $\mathrm{i}$ cross section
is the cross section for events that contain one or more~$\mathrm{i}$,
\begin{equation}
 \label{eq:altIncl}
  \sigma_{\ge1\mathrm{i}} \equiv \sum_{n=1} \sigma_{n\mathrm{i}}\,.
\end{equation}
In the eikonal model, this is given by
\begin{equation}
  \hspace*{-1em}
  \sigma_{\ge1\mathrm{i}}
  = \sum_{n=1} \int \mathrm{d}^2b \,
  \frac{\left(\sigma_{\mathrm{i}}A(b)\right)^n}{n!} \,
  \mathrm{e}^{-\sigma_{\mathrm{i}}A(b)}
  = \int \mathrm{d}^2b \left[\sum_{n=1}
  \frac{\left(\sigma_{\mathrm{i}}A(b)\right)^n}{n!}\right]
  \mathrm{e}^{-\sigma_{\mathrm{i}}A(b)}
  = \int \mathrm{d}^2b\left(1-\mathrm{e}^{-\sigma_{\mathrm{i}}A(b)}\right).
  \hspace*{-1em}
\end{equation}

\subsubsection{Exclusive definition}

The exclusive single-$i$ cross section has already been defined,
$\sigma_{1\mathrm{i}}=\sigma_{n\mathrm{i}}$, with $n=1$, which in the
eikonal model is given by
\begin{equation}
  \sigma_{1\mathrm{i}}
  = \int \mathrm{d}^2b \,
  \sigma_{\mathrm{i}}A(b) \, \mathrm{e}^{-\sigma_{\mathrm{i}}A(b)}\,.
\end{equation}

\subsection{Double-particle cross sections}

\subsubsection{Inclusive definition}
As we again motivate in more detail in
Appendix~\ref{sect:inclusiveCrossSectionsDetails}, the double-particle
inclusive cross section is given by
\begin{equation}
  \sigma_{\textit{incl\/}\mathrm{ii}}
  = \sum_n \mbox{$\frac12$}n(n\!-\!1)\sigma_{n\mathrm{i}}\,.
\end{equation}
An event that contains $n$ $\mathrm{i}$'s contains $\frac12n(n\!-\!1)$
different $\mathrm{ii}$ pairs and contributes that many times to the
inclusive cross section.

It is worth noting that this definition imposes a specific requirement
on the experimental measurement. Since the reconstruction efficiency is
typically small, but not infinitesimal, it can happen that more than
two $\mathrm{i}$'s are reconstructed in the same event. The inclusive
cross section definition requires that in such events each pair
contributes to the cross section. Thus, a three-$\mathrm{i}$ event
contributes three times to the double-$\mathrm{i}$ cross section. The
event sample used for LHCb's measurement is not large enough for this to
be an issue\cite{Vanya}, but it could be in future.

In the eikonal model, the double-particle inclusive cross section is
therefore given by
\begin{equation}
  \sigma_{\textit{incl\/}\mathrm{ii}} = \sum_n \mbox{$\frac12$}n(n\!-\!1)\int \mathrm{d}^2b \,
  \frac{\left(\sigma_{\mathrm{i}}A(b)\right)^n}{n!} \,
  \mathrm{e}^{-\sigma_{\mathrm{i}}A(b)}
  = \frac12\sigma_{\mathrm{i}}^2\int \mathrm{d}^2b \, A(b)^2\,.
\end{equation}

In a completely analogous way, the double-particle cross sections for two
different particle types $\mathrm{i}$ and $\mathrm{j}$ can be defined in
terms of the exclusive cross section for events containing $n$
$\mathrm{i}$'s and $m$ $\mathrm{j}$'s as
\begin{equation}
  \sigma_{\textit{incl\/}\mathrm{ij}}
  = \sum_{n,m} n\,m\,\sigma_{n\mathrm{i},m\mathrm{j}}\,.
\end{equation}
In the eikonal model this gives
\begin{equation}
  \sigma_{\textit{incl\/}\mathrm{ij}}
  = \sigma_{\mathrm{i}}\,\sigma_{\mathrm{j}}\int \mathrm{d}^2b \, A(b)^2\,.
\end{equation}

\subsubsection{Alternative inclusive definition}

The alternative inclusive definition is the cross section for events
containing two or more~$\mathrm{i}$'s,
\begin{equation}
  \sigma_{\ge2\mathrm{i}} \equiv \sum_{n=2} \sigma_{n\mathrm{i}}\,.
\end{equation}
In the eikonal model, this is given by
\begin{equation}
  \sigma_{\ge2\mathrm{i}}
  = \int \mathrm{d}^2b\left(1-\mathrm{e}^{-\sigma_{\mathrm{i}}A(b)}
  -\sigma_{\mathrm{i}}A(b)\mathrm{e}^{-\sigma_{\mathrm{i}}A(b)}\right).
\end{equation}
By analogy, we have
\begin{equation}
  \sigma_{\ge1\mathrm{i},\ge1\mathrm{j}} \equiv \sum_{n,m=1} \sigma_{n\mathrm{i},m\mathrm{j}}\,,
\end{equation}
and
\begin{equation}
  \sigma_{\ge1\mathrm{i},\ge1\mathrm{j}}
  = \int
  \mathrm{d}^2b\left(1-\mathrm{e}^{-\sigma_{\mathrm{i}}A(b)}\right)
  \left(1-\mathrm{e}^{-\sigma_{\mathrm{j}}A(b)}\right).
\end{equation}

\subsubsection{Exclusive definition}

The exclusive double-$i$ cross section has already been defined, and in
the eikonal model is given by
\begin{equation}
  \sigma_{2\mathrm{i}}
  = \int \mathrm{d}^2b \,
  \mbox{$\frac12$}(\sigma_{\mathrm{i}}A(b))^2 \,
  \mathrm{e}^{-\sigma_{\mathrm{i}}A(b)}\,,
\end{equation}
and
\begin{equation}
  \sigma_{1\mathrm{i},1\mathrm{j}}
  = \int
  \mathrm{d}^2b\,\sigma_{\mathrm{i}}\sigma_{\mathrm{j}}\,A(b)^2\,
  \mathrm{e}^{-(\sigma_{\mathrm{i}}+\sigma_{\mathrm{j}})A(b)}\,.
\end{equation}

\subsection{The effective cross section}

In all cases, we take Eq.~(\ref{eq:sigmaeffdef}) as the definition of
\seff.

\subsubsection{Inclusive definition}

\begin{equation}
  \seff =
  \frac{\sigma_{\textit{incl\/}\mathrm{i}}^2}{2\sigma_{\textit{incl\/}\mathrm{ii}}}
  =\frac{\sigma_{\textit{incl\/}\mathrm{i}}\,\sigma_{\textit{incl\/}\mathrm{j}}}
  {\sigma_{\textit{incl\/}\mathrm{ij}}}
  = \frac1{\int \mathrm{d}^2b \, A(b)^2}\,.
\end{equation}
Note that in the eikonal model, the effective cross section is
independent of the cross section $\sigma_{\mathrm{i}}$, and hence the
choice of $\mathrm{i}$.

\subsubsection{Alternative inclusive definition}

\begin{equation}
  \seff =
  \frac{\sigma_{\ge1\mathrm{i}}^2}{2\sigma_{\ge2\mathrm{i}}}
  = \frac{\left[\int \mathrm{d}^2b\left(1-\mathrm{e}^{-\sigma_{\mathrm{i}}A(b)}\right)\right]^2}
  {2\int \mathrm{d}^2b\left(1-\mathrm{e}^{-\sigma_{\mathrm{i}}A(b)}
  -\sigma_{\mathrm{i}}A(b)\mathrm{e}^{-\sigma_{\mathrm{i}}A(b)}\right)}\,.
\end{equation}
With this definition, \seff is process-dependent, since its value
depends on $\sigma_{\mathrm{i}}$. The two different definitions in
Eq.~(\ref{eq:sigmaeffdef}) give different results,
\begin{equation}
  \seff =
  \frac{\sigma_{\ge1\mathrm{i}}\sigma_{\ge1\mathrm{j}}}{\sigma_{\ge1\mathrm{i},\ge1\mathrm{j}}}
  = \frac{
    \left[\int \mathrm{d}^2b\left(1-\mathrm{e}^{-\sigma_{\mathrm{i}}A(b)}\right)\right]
    \left[\int \mathrm{d}^2b\left(1-\mathrm{e}^{-\sigma_{\mathrm{j}}A(b)}\right)\right]
    }{\int
  \mathrm{d}^2b\left(1-\mathrm{e}^{-\sigma_{\mathrm{i}}A(b)}\right)
  \left(1-\mathrm{e}^{-\sigma_{\mathrm{j}}A(b)}\right)}\,.
\end{equation}

\subsubsection{Exclusive definition}

\begin{equation}
  \seff =
  \frac{\sigma_{1\mathrm{i}}^2}{2\sigma_{2\mathrm{i}}}
  = \frac{\left[\int \mathrm{d}^2b \,
  A(b) \, \mathrm{e}^{-\sigma_{\mathrm{i}}A(b)}\right]^2}
  {\int \mathrm{d}^2b \,
  A(b)^2 \,
  \mathrm{e}^{-\sigma_{\mathrm{i}}A(b)}}\,.
\end{equation}
\begin{equation}
  \seff =
  \frac{\sigma_{1\mathrm{i}}\sigma_{1\mathrm{j}}}{\sigma_{1\mathrm{i},1\mathrm{j}}}
  = \frac{
    \left[\int \mathrm{d}^2b \, A(b) \, \mathrm{e}^{-\sigma_{\mathrm{i}}A(b)}\right]
    \left[\int \mathrm{d}^2b \, A(b) \, \mathrm{e}^{-\sigma_{\mathrm{j}}A(b)}\right]
    }{\int
  \mathrm{d}^2b\,A(b)^2\,
  \mathrm{e}^{-(\sigma_{\mathrm{i}}+\sigma_{\mathrm{j}})A(b)}}\,.
\end{equation}

\subsubsection{Conclusion}

The conclusion of this study is that, of the different cross section
definitions used in the literature, only the conventional inclusive one
gives an expression for \seff that is independent of the chosen
process(es), in the eikonal model. That is, for the single-particle
inclusive $\mathrm{i}$ cross section, particles of type $\mathrm{i}$
should be counted, so that an event that contains $n$ $\mathrm{i}$'s
contributes $n$ times. For the double-particle inclusive $\mathrm{ii}$
cross section, pairs of particles of type $\mathrm{i}$ should be
counted, so that an event that contains $n$ $\mathrm{i}$'s contributes
$\frac12n(n\!-\!1)$ times. For the double-particle inclusive
$\mathrm{ij}$ cross section, pairs of particles of type $\mathrm{i}$ and
$\mathrm{j}$ should be counted, so that an event that contains $n$
$\mathrm{i}$'s and $m$ $\mathrm{j}$'s contributes $nm$ times. We use
these definitions for the remainder of the paper and drop the subscript
$\textit{incl\/}$ from them.

\section{Charm quark and charmed hadron cross sections}
\label{sect:charmCrossSections}

The discussion of the previous section can be applied directly to the
partonic cross sections to produce charm quark pairs. We include all
pair production mechanisms, whether directly through
$\mathrm{q\bar{q}\to c\bar{c}}$ and $\mathrm{gg\to c\bar{c}}$ or through
flavour excitation, e.g.\ $\mathrm{qc\to qc}$, with an accompanying
$\mathrm{\bar{c}}$ produced in the corresponding initial state
shower. For simplicity we neglect the possibility that a single partonic
collision produces more than one $\mathrm{c\bar{c}}$ pair.

However, experiments do not directly observe charm quarks, but rather
the hadrons they fragment to, whether charmonium,
e.g.\ $\mathrm{J}/\psi$, or open charm, e.g.\ the set of $\mathrm{D}$
mesons or $\Lambda_{\mathrm{c}}$ baryons. We concentrate on the case of
open charm.

\subsection{Open charm cross sections}

In QCD, the production of charmed hadrons can be factorized into a hard
process, which produces a charm quark, and a perturbative evolution
followed by the non-perturbative confinement of the charm quark into a
charmed hadron. The last two processes are collectively called
fragmentation. In the present discussion we neglect the possibility
that the evolution of a gluon or quark could produce a
charm-anticharm pair and hence the fragmentation process preserves the
charm quantum number: a charm quark produces a charmed hadron with unit
probability. We further assume that the hadronization stage is a local
process and, hence, the probability distributions of which charmed
hadron a given charm quark produces are independent.

The fragmentation process typically degrades the energy of the charm
quark so that the produced charmed hadron has less energy than it
(although not necessarily in the laboratory frame, a point that we shall
return to in the next sub-section), without a significant change in
direction. Thus, the kinematic distributions of the produced charmed
hadrons are related to those of the initiating charm quarks, but folded
with fragmentation functions. For the present analysis, we will assume
that the kinematic distributions of $\mathrm{c\bar{c}}$ pairs produced
in different partonic collisions are independent. Thus, the probability
that a given charm quark produces a charmed hadron of a given species
within the fiducial region of an experiment is a fixed number,
$p^{\mathrm{c}}_{\mathrm{D}}$. (We use a generic $\mathrm{D}$ label for
a charmed hadron, although it could also be a charmed
baryon. Specifically, we are interested in the cases
$\mathrm{D}=\{\mathrm{D^0},\mathrm{D^+},\mathrm{D^+_s},\Lambda_{\mathrm{c}}\}$).
We assume that $p^{\mathrm{\bar{c}}}_{\mathrm{D}}=0$.

Within these assumptions, and using the single- and double-inclusive
charm quark cross sections, it is straightforward to calculate the
single- and double-inclusive charmed hadron cross sections. An event
containing $n$ charm quarks has an independent probability
$p^{\mathrm{c}}_{\mathrm{D}}$ for each of them to produce a $\mathrm{D}$
hadron and hence
\begin{eqnarray}
  \sigma_{\mathrm{D}} &=& p^{\mathrm{c}}_{\mathrm{D}}\sigma_{\mathrm{c}}\,, \\
  \sigma_{\mathrm{DD}} &=& (p^{\mathrm{c}}_{\mathrm{D}})^2\sigma_{\mathrm{cc}}
\end{eqnarray}
(it is worth noting that these relations are not true for the
alternative inclusive or exclusive cross section definitions). Likewise,
a pair of charmed hadrons of different species, $\mathrm{D}_1$ and
$\mathrm{D}_2$, can be produced by a pair of charm quarks in either of
two ways, and we have
\begin{equation}
  \sigma_{\mathrm{D_1D_2}} = 2p^{\mathrm{c}}_{\mathrm{D_1}}p^{\mathrm{c}}_{\mathrm{D_2}}
    \sigma_{\mathrm{cc}}\,.
\end{equation}
Thus, we can use measurements of single- and double-charmed hadron
production to extract \seff, independent of the unknown charm quark
cross section and fragmentation probabilities:
\begin{eqnarray}
  \frac{\sigma_{\mathrm{D}}^2}{2\sigma_{\mathrm{DD}}} \;=&\displaystyle
  \frac{(p^{\mathrm{c}}_{\mathrm{D}})^2\sigma_{\mathrm{c}}^2}
       {2(p^{\mathrm{c}}_{\mathrm{D}})^2\sigma_{\mathrm{cc}}} &=\;
  \seff\,, \\
  \frac{\sigma_{\mathrm{D_1}}\sigma_{\mathrm{D_2}}}{\sigma_{\mathrm{D_1D_2}}} \;=&\displaystyle
  \frac{p^{\mathrm{c}}_{\mathrm{D_1}}p^{\mathrm{c}}_{\mathrm{D_2}}\sigma_{\mathrm{c}}^2}
       {2p^{\mathrm{c}}_{\mathrm{D_1}}p^{\mathrm{c}}_{\mathrm{D_2}}\sigma_{\mathrm{cc}}} &=\;
  \seff\,.
\end{eqnarray}

\subsection{Charge conjugate modes}

Since QCD is charge-conjugation-symmetric, one might expect that
$p^{\mathrm{c}}_{\mathrm{D}}=p^{\mathrm{\bar{c}}}_{\mathrm{\overline{D}}}$. However,
the fact that the LHC collides particles, rather than antiparticles, can
in principle induce an asymmetry. One expects that the primary
production distributions of charm and anti\-charm are the same and
likewise, because it is local, the probability distribution of which
charmed hadron is produced. However, the kinematic distributions of
produced charmed and anticharmed hadrons are not necessarily the same,
because the colour structure of their production is different. The
colour partner of a charm quark is more likely to be the proton remnant,
or lie towards the proton remnant direction, whereas the colour partner
of an anticharm quark is more likely to be in the final state of the
hard process, and therefore towards the centre of the event. The
hadronization phase is more properly thought of as being largely
longitudinal \emph{in the rest frame of colour-connected pairs} and
hence, on average, charmed hadrons are expected to be produced at
slightly higher rapidities than their parent charm quarks, while
anticharmed hadrons are expected to be produced at the same rapidity as
their parent anticharm quarks. This effect was called ``string drag'' in
Ref.~\cite{Norrbin:1998bw,Norrbin:2000zc}, but as it occurs in other
hadronization
models, we prefer to call it a ``colour drag''. Thus, while the total
number of $\mathrm{D}$ hadrons of a given species is expected to be the
same as the number of $\mathrm{\overline{D}}$ hadrons, the numbers
within a given fiducial volume are not necessarily the same.

In the analysis we are comparing to,
LHCb did not notice any difference between charge conjugate modes and
hence did not explicitly extract a measurement for this
asymmetry\cite{Vanya}.
They have made dedicated analyses of $\mathrm{D_s^\pm}$ production
asymmetries\cite{LHCb:2012fb} and found no effect (at the $<1\%$ level)
and $\mathrm{D}^\pm$\cite{Aaij:2012cy} and found $3\sigma$ evidence for
an asymmetry at the $\sim1\%$ level. It would be interesting to take
this effect into account explicitly in any future measurement, but it
appears to be small enough that in the remainder, we follow
LHCb\cite{Aaij:2012dz} in assuming
\begin{equation}
  \label{eq:assumption}
  p^{\mathrm{c}}_{\mathrm{D}}=p^{\mathrm{\bar{c}}}_{\mathrm{\overline{D}}}\,,
\end{equation}
and therefore that the cross sections for $\mathrm{D}$ and
$\mathrm{\overline{D}}$ are equal,
\begin{eqnarray}
  \sigma_{\mathrm{D}} &=& \sigma_{\mathrm{\overline{D}}}\,, \\
  \sigma_{\mathrm{D}\mathrm{D}} &=& \sigma_{\mathrm{\overline{D}}\,\mathrm{\overline{D}}}\,, \\
  \sigma_{\mathrm{D_1}\mathrm{D_2}} &=& \sigma_{\mathrm{\overline{D}_1}\mathrm{\overline{D}_2}}\,.
\end{eqnarray}
LHCb used these results to effectively double their data set and defined
\begin{eqnarray}
  \sigma_{\mathrm{D,LHCb}} &\equiv& \sigma_{\mathrm{D}} + \sigma_{\mathrm{\overline{D}}}\,, \\
  \sigma_{\mathrm{D}\mathrm{D,LHCb}} &\equiv& \sigma_{\mathrm{D}\mathrm{D}} + \sigma_{\mathrm{\overline{D}}\,\mathrm{\overline{D}}}\,, \\
  \sigma_{\mathrm{D_1}\mathrm{D_2,LHCb}} &\equiv& \sigma_{\mathrm{D_1}\mathrm{D_2}} + \sigma_{\mathrm{\overline{D}_1}\mathrm{\overline{D}_2}}\,.
\end{eqnarray}
It is still possible to use these charge-conjugation-summed cross
sections to extract \seff, at least under the assumption
(\ref{eq:assumption}), but one must be careful to include an additional
factor of~2:
\begin{eqnarray}
  \frac{\sigma_{\mathrm{D,LHCb}}^2}{\mbox{\boldmath$2\times$}2\sigma_{\mathrm{DD,LHCb}}} =
  \frac{(\sigma_{\mathrm{D}}+\sigma_{\mathrm{\overline{D}}})^2}{4(\sigma_{\mathrm{D}\mathrm{D}} + \sigma_{\mathrm{\overline{D}}\,\mathrm{\overline{D}}})} =
  \frac{4\sigma_{\mathrm{D}}^2}{8\sigma_{\mathrm{D}\mathrm{D}}}
  &=&
  \seff\,, \\
  \frac{\sigma_{\mathrm{D_1,LHCb}}\sigma_{\mathrm{D_2,LHCb}}}{\mbox{\boldmath$2\times$}\sigma_{\mathrm{D_1D_2,LHCb}}} =
  \frac{(\sigma_{\mathrm{D_1}}+\sigma_{\mathrm{\overline{D}_1}})(\sigma_{\mathrm{D_2}}+\sigma_{\mathrm{\overline{D}_2}})}
  {2(\sigma_{\mathrm{D_1}\mathrm{D_2}} + \sigma_{\mathrm{\overline{D}_1}\mathrm{\overline{D}_2}})} =
  \frac{4\sigma_{\mathrm{D_1}}\sigma_{\mathrm{D_2}}}
  {4\sigma_{\mathrm{D_1}\mathrm{D_2}}}
  &=&
  \seff\,.
\end{eqnarray}
It appears to us that LHCb have not included this factor of two and
hence that their extracted values of \seff are too large by a factor of
two.

It is worth mentioning that charmonium channels are not subject to this
factor of two. Since charmonium is self-conjugate, there is no summation
to be done. The double-charmonium channel therefore does not contain any
additional factors of two. The single-charmonium, single-open charm
channel contains a factor of two in the numerator, from the sum over
$\mathrm{D}$ and $\mathrm{\overline{D}}$, but also in the denominator,
from the sum over $\mathrm{J/\psi+D}$ and
$\mathrm{J/\psi+\overline{D}}$. Thus the two factors of two cancel.

We summarize LHCb's results, corrected by this factor of~2 in
Sect.~\ref{sect:results}.

\subsection{Opposite-sign charmed hadron pairs}

While the main focus of this paper is double- (and single-) inclusive
production of charmed hadrons both containing a charm quark (or both an
anticharm), we briefly mention the channels in which a charmed and
anticharmed hadron pair are detected, which have also been measured by
LHCb. Even within the assumption
$p^{\mathrm{\bar{c}}}_{\mathrm{D}}=p^{\mathrm{c}}_{\mathrm{\overline{D}}}=0$,
$\mathrm{D\overline{D}}$ and $\mathrm{D_1\overline{D}_2}$ pairs can come
from a single $\mathrm{c\bar{c}}$ pair, which have equal and opposite
transverse momenta and correlated rapidities. Therefore we cannot
consider the probabilities of charm quarks to produce charmed hadrons
within the fiducial region as uncorrelated\footnote{Rapidity
  correlations were proposed as a means to separate single- and
  double-charm production in Ref.~\cite{Gaunt:2011rm}.}.

We introduce a correlation coefficient $\mathcal{C}$, such that the
probabilities that a $\mathrm{c\bar{c}}$ pair from a single partonic
scattering produces a $\mathrm{D\overline{D}}$ or
$\mathrm{D_1\overline{D}_2}$ pair within the fiducial region are
$\mathcal{C}(p^{\mathrm{c}}_{\mathrm{D}})^2$ and
$\mathcal{C}p^{\mathrm{c}}_{\mathrm{D_1}}p^{\mathrm{c}}_{\mathrm{D_2}}$
respectively. The corresponding probabilities that a $\mathrm{c}$ and a
$\mathrm{\bar{c}}$ from different partonic scatterings produces a
$\mathrm{D\overline{D}}$ or $\mathrm{D_1\overline{D}_2}$ pair within the
fiducial region are still $(p^{\mathrm{c}}_{\mathrm{D}})^2$ and
$p^{\mathrm{c}}_{\mathrm{D_1}}p^{\mathrm{c}}_{\mathrm{D_2}}$
respectively.

We can then show that
\begin{eqnarray}
  \sigma_{\mathrm{D\overline{D}}} &=& (p^{\mathrm{c}}_{\mathrm{D}})^2
  \bigl(\mathcal{C}\sigma_{\mathrm{c}}+2\sigma_{\mathrm{cc}}\bigr), \\
  \sigma_{\mathrm{D_1\overline{D}_2}} &=& p^{\mathrm{c}}_{\mathrm{D_1}}p^{\mathrm{c}}_{\mathrm{D_2}}
  \bigl(\mathcal{C}\sigma_{\mathrm{c}}+2\sigma_{\mathrm{cc}}\bigr).
\end{eqnarray}
Then, forming the same ratios as in the like-sign case, we obtain:
\begin{eqnarray}
  \label{eq:LHCbdef1}
  \frac{\sigma_{\mathrm{D}}^2}{2\sigma_{\mathrm{D\overline{D}}}} &=&
  \frac{\seff}{2(\mathcal{C}\seff/\sigma_{\mathrm{c}}+1)}\,, \\
  \label{eq:LHCbdef2}
  \frac{\sigma_{\mathrm{D_1}}\sigma_{\mathrm{D_2}}}{\sigma_{\mathrm{D_1\overline{D}_2}}} &=&
  \frac{\seff}{\mathcal{C}\seff/\sigma_{\mathrm{c}}+1}\,.
\end{eqnarray}
Note that both $\mathcal{C}$ and $\seff/\sigma_{\mathrm{c}}$ are
expected to be larger than 1. Thus, these results are expected to be
significantly smaller than \seff, but without further studies to extract
the values of these constants, we cannot quantify the expected size. It
is important to note however that they are independent of the specific
flavours of the charmed hadrons.

Finally, we note again that LHCb sum over charge conjugate modes,
\begin{eqnarray}
  \sigma_{\mathrm{D}\mathrm{\overline{D},LHCb}} &\equiv& \sigma_{\mathrm{D}\mathrm{\overline{D}}}\,, \\
  \sigma_{\mathrm{D_1}\mathrm{\overline{D}_2,LHCb}} &\equiv& \sigma_{\mathrm{D_1}\mathrm{\overline{D}_2}} + \sigma_{\mathrm{\overline{D}_1}\mathrm{D_2}}\,.
\end{eqnarray}
Therefore when they form the ratios in Eqs.~(\ref{eq:LHCbdef1})
and~(\ref{eq:LHCbdef2}), they obtain
\begin{eqnarray}
  \frac{\sigma_{\mathrm{D,LHCb}}^2}{2\sigma_{\mathrm{D\overline{D},LHCb}}}
  \;=&\displaystyle
  \frac{4\sigma_{\mathrm{D}}^2}{2\sigma_{\mathrm{D\overline{D}}}} &=\;
  \frac{2\seff}{\mathcal{C}\seff/\sigma_{\mathrm{c}}+1}\,, \\
  \frac{\sigma_{\mathrm{D_1,LHCb}}\sigma_{\mathrm{D_2,LHCb}}}{\sigma_{\mathrm{D_1\overline{D}_2,LHCb}}}
  \;=&\displaystyle
  \frac{4\sigma_{\mathrm{D_1}}\sigma_{\mathrm{D_2}}}{2\sigma_{\mathrm{D_1\overline{D}_2,LHCb}}} &=\;
  \frac{2\seff}{\mathcal{C}\seff/\sigma_{\mathrm{c}}+1}\,.
\end{eqnarray}
That is, defined in this way, the ratios are flavour-independent, as
LHCb noted, and still smaller than \seff. Since we do not have reliable
estimates for $\mathcal{C}$ and $\seff/\sigma_{\mathrm{c}}$, we do not
consider these opposite-sign cases further.

\section{Results}
\label{sect:results}

\begin{figure*}[htb]
  \begin{center}
\includegraphics[angle=90,width=0.75\textwidth]{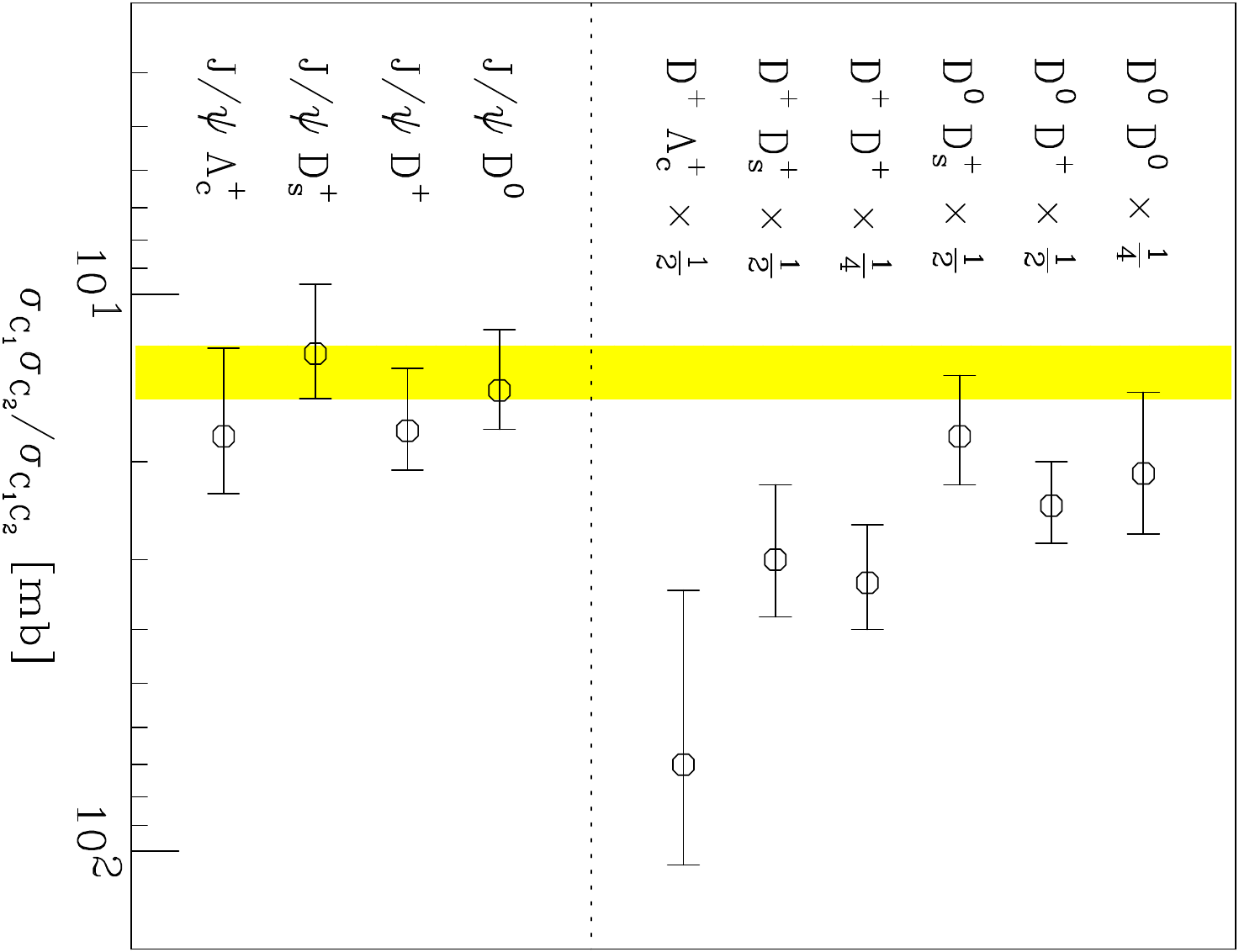}
  \end{center}
  \caption{The values of \seff extracted from various LHCb measurements
    (data points, with statistical and systematic errors added in
    quadrature) compared with the value extracted from CDF and D0 data
    (yellow band).}
  \label{fig:seff}
\end{figure*}

Our main result is the fact that the LHCb extraction of \seff from their
double-inclusive open-charm data is too large by a factor of~2. Applying
this factor of~2, we summarize in Fig.~\ref{fig:seff} their results both
for the open-charm channels and the channels with a $\mathrm{J}/\psi$,
together with the average of the D0 and corrected\cite{Bahr:2013gkj} CDF
results quoted in Ref.~\cite{Seymour:2013qka}.

It is clear that while the additional factor of~2 has brought the
results closer together, the results for double open charm are still
significantly higher than for the other processes.

\optional{\section{Monte Carlo study of double-charm production}
\label{sect:MonteCarlo}}

\section{Summary}
\label{sect:summary}

We have discussed various issues related to the measurement of
double-charm cross sections and the extraction of the effective cross
section for double-parton scattering, \seff, from them. We have
emphasized the importance of properly-defined inclusive cross sections,
in which each final state particle, or particle pair, of the given type
is counted. With this definition, the effective cross section can be
extracted directly from charmed hadron data without needing further
information from theory, and is also a process-independent quantity in
the commonly-used eikonal model.

We have noticed that LHCb use both single- and double-open charm cross
section definitions that are summed over charge conjugate modes and
that, therefore, an additional factor of~2 needs to be applied to the
extraction of \seff from these cross sections. This brings the data from
open-charm pairs closer to that from $\mathrm{J}/\psi$ plus open charm
and jet production, but it still lies considerably higher.

We have mentioned several issues that could be worthy of further study:
triple-charm production; differences between charmed and anticharmed
hadron distributions; and correlations between charm-anticharm pairs.

In Appendix~\ref{sect:Berezhnoy} below, we make comparisons with another
theory paper that has compared with the same LHCb measurements.

\section*{Acknowledgements}

We are grateful to Vanya Belyaev, Marco Gersabeck and Chris Parkes for
discussions of the LHCb analysis. This work was funded in part by the
Lancaster-Manchester-Sheffield Consortium for Fundamental Physics under
STFC grant ST/J000418/1 and in part by the MCnetITN FP7 Marie Curie
Initial Training Network PITN-GA-2012-315877.

\appendix

\section{Cross section definitions in more detail}
\label{sect:inclusiveCrossSectionsDetails}

In this Appendix we prove some of the results quoted in
Sect.~\ref{sect:inclusiveCrossSections} and illustrate their discussion
with their approximations in the limit of small cross sections. There is
inevitably some overlap with the discussion of
Sect.~\ref{sect:inclusiveCrossSections}, but we try to keep this to a
minimum, resulting in a rather terse presentation. Nevertheless, we hope
that interested readers will be able to follow this discussion and
reconstruct our argument if necessary.

\subsection{Single-particle cross sections}

\subsubsection{Inclusive definition}

The number of $\mathrm{i}$'s counted, $N_{\mathrm{i}}$, in a run with
integrated luminosity $\mathcal{L}$ defines the inclusive $\mathrm{i}$
cross section:
\begin{equation}
  \sigma_{\textit{incl\/}\mathrm{i}} \equiv
  \frac{N_{\mathrm{i}}}{\mathcal{L}}\,.
\end{equation}
There is an alternative way of writing this cross section that will help
us in generalizing to multi-particle cross sections. As a first step,
one can imagine that our hypothetical counter has a vanishingly small
reconstruction efficiency $\epsilon_{rec}$, but that the reconstruction
probability of a
given $\mathrm{i}$ is independent of any other $\mathrm{i}$'s in the
event. Thus the inclusive cross section is now written as
\begin{equation}
  \sigma_{\textit{incl\/}\mathrm{i}} =
  \lim_{\epsilon_{\textit{rec}}\to0}
  \frac{N_{\mathrm{i}\textit{rec}}}{\epsilon_{\textit{rec}}\mathcal{L}}\,.
\end{equation}
In this limit, it never happens that more than one $\mathrm{i}$ is
reconstructed in the same event. In an event in which there are $n$
$\mathrm{i}$'s, the probability that one of them is reconstructed is
$n\epsilon_{\textit{rec}}$. We therefore have
\begin{eqnarray}
  \sigma_{\textit{incl\/}\mathrm{i}}
  &=&
  \lim_{\epsilon_{\textit{rec}}\to0}
  \sum_n
  \frac{n\epsilon_{\textit{rec}}N_{n\mathrm{i}}}
       {\epsilon_{\textit{rec}}\mathcal{L}} \\
  &=& \sum_n
  \frac{nN_{n\mathrm{i}}}
       {\mathcal{L}}\,,
\end{eqnarray}
where $N_{n\mathrm{i}}$ is the number of events in which there are
exactly $n$ $\mathrm{i}$'s. Finally, we define the exclusive cross
section for events in which there are exactly $n$ $\mathrm{i}$'s,
\begin{equation}
  \sigma_{n\mathrm{i}} \equiv \frac{N_{n\mathrm{i}}}{\mathcal{L}}\,,
\end{equation}
and hence
\begin{equation}
  \sigma_{\textit{incl\/}\mathrm{i}} = \sum_n n\,\sigma_{n\mathrm{i}}\,.
\end{equation}
That is, an event that contains $n$ $\mathrm{i}$'s contributes $n$ times
to the inclusive cross section.

The inclusive cross section in the eikonal model is then
\begin{eqnarray}
  \sigma_{\textit{incl\/}\mathrm{i}} &=& \sum_n n\,\int \mathrm{d}^2b \,
  \frac{\left(\sigma_{\mathrm{i}}A(b)\right)^n}{n!} \,
  \mathrm{e}^{-\sigma_{\mathrm{i}}A(b)}
  = \int \mathrm{d}^2b \left[\sum_n n
  \frac{\left(\sigma_{\mathrm{i}}A(b)\right)^n}{n!}\right]
  \mathrm{e}^{-\sigma_{\mathrm{i}}A(b)}
  = \int \mathrm{d}^2b \, \sigma_{\mathrm{i}}A(b)
\nonumber\\
  &=& \sigma_{\mathrm{i}}\,.
\end{eqnarray}

\subsubsection{Alternative inclusive definition}

In the alternative inclusive definition, the inclusive $\mathrm{i}$
cross section is the cross section for events that contain one or
more~$\mathrm{i}$,
\begin{equation}
  \sigma_{\ge1\mathrm{i}} \equiv \sum_{n=1} \sigma_{n\mathrm{i}}\,.
\end{equation}
In the eikonal model, this is given by
\begin{equation}
  \hspace*{-1em}
  \sigma_{\ge1\mathrm{i}}
  = \sum_{n=1} \int \mathrm{d}^2b \,
  \frac{\left(\sigma_{\mathrm{i}}A(b)\right)^n}{n!} \,
  \mathrm{e}^{-\sigma_{\mathrm{i}}A(b)}
  = \int \mathrm{d}^2b \left[\sum_{n=1}
  \frac{\left(\sigma_{\mathrm{i}}A(b)\right)^n}{n!}\right]
  \mathrm{e}^{-\sigma_{\mathrm{i}}A(b)}
  = \int \mathrm{d}^2b\left(1-\mathrm{e}^{-\sigma_{\mathrm{i}}A(b)}\right).
  \hspace*{-1em}
\end{equation}
For small $\sigma_{\mathrm{i}}$ this agrees with the conventional
definition, but for larger $\sigma_{\mathrm{i}}$ it clearly differs,
\begin{equation}
  \sigma_{\ge1\mathrm{i}}
  \approx \sigma_{\mathrm{i}}
  -\frac12\sigma_{\mathrm{i}}^2
  \int \mathrm{d}^2b \, A(b)^2
  +\mathcal{O}(\sigma_{\mathrm{i}}^3).
\end{equation}

\subsubsection{Exclusive definition}

\begin{equation}
  \sigma_{1\mathrm{i}}
  = \int \mathrm{d}^2b \,
  \sigma_{\mathrm{i}}A(b) \, \mathrm{e}^{-\sigma_{\mathrm{i}}A(b)}\,.
\end{equation}
Again, for small $\sigma_{\mathrm{i}}$ this agrees with the inclusive
definition, but for larger $\sigma_{\mathrm{i}}$ it differs,
\begin{equation}
  \sigma_{1\mathrm{i}}
  \approx \sigma_{\mathrm{i}}
  -\sigma_{\mathrm{i}}^2
  \int \mathrm{d}^2b \, A(b)^2
  +\mathcal{O}(\sigma_{\mathrm{i}}^3).
\end{equation}

\subsection{Double-particle cross sections}

\subsubsection{Inclusive definition}
For the formal definition of the two-particle inclusive cross section,
we return to our hypothetical particle counter. We count the number of
times in which it counts two $\mathrm{i}$'s in the same event:
\begin{equation}
  \sigma_{\textit{incl\/}\mathrm{ii}}
  =
  \lim_{\epsilon_{\textit{rec}}\to0}
  \frac{N_{\mathrm{ii}\textit{rec}}}
       {\epsilon_{\textit{rec}}^2\mathcal{L}}\,.
\end{equation}
In this limit, it never happens that more than two $\mathrm{i}$'s are
reconstructed in the same event. In an event in which there are $n$
$\mathrm{i}$'s, the probability that two of them are reconstructed is
given by the binomial probability
$\left(n\atop2\right)\epsilon_{\textit{rec}}^2 =
\frac12n(n\!-\!1)\epsilon_{\textit{rec}}^2$. We therefore have
\begin{equation}
  \sigma_{\textit{incl\/}\mathrm{ii}}
  = \sum_n \mbox{$\frac12$}n(n\!-\!1)\sigma_{n\mathrm{i}}\,.
\end{equation}
That is, an event that contains $n$ $\mathrm{i}$'s contains
$\frac12n(n\!-\!1)$ different $\mathrm{ii}$ pairs and hence contributes
that many times to the inclusive cross section.

In the eikonal model, the double-particle inclusive cross section is
given by
\begin{equation}
  \sigma_{\textit{incl\/}\mathrm{ii}} = \sum_n \mbox{$\frac12$}n(n\!-\!1)\int \mathrm{d}^2b \,
  \frac{\left(\sigma_{\mathrm{i}}A(b)\right)^n}{n!} \,
  \mathrm{e}^{-\sigma_{\mathrm{i}}A(b)}
  = \frac12\sigma_{\mathrm{i}}^2\int \mathrm{d}^2b \, A(b)^2\,.
\end{equation}
For two different particle types $\mathrm{i}$ and $\mathrm{j}$,
\begin{equation}
  \sigma_{\textit{incl\/}\mathrm{ij}}
  = \sum_{n,m} n\,m\,\sigma_{n\mathrm{i},m\mathrm{j}}\,,
\end{equation}
\begin{equation}
  \sigma_{\textit{incl\/}\mathrm{ij}}
  = \sigma_{\mathrm{i}}\,\sigma_{\mathrm{j}}\int \mathrm{d}^2b \, A(b)^2\,.
\end{equation}

\subsubsection{Alternative inclusive definition}

\begin{equation}
  \sigma_{\ge2\mathrm{i}} \equiv \sum_{n=2} \sigma_{n\mathrm{i}}\,,
\end{equation}
\begin{equation}
  \sigma_{\ge2\mathrm{i}}
  = \int \mathrm{d}^2b\left(1-\mathrm{e}^{-\sigma_{\mathrm{i}}A(b)}
  -\sigma_{\mathrm{i}}A(b)\mathrm{e}^{-\sigma_{\mathrm{i}}A(b)}\right).
\end{equation}
For small $\sigma_{\mathrm{i}}$ this agrees with the conventional
definition, but for larger $\sigma_{\mathrm{i}}$ it clearly differs,
\begin{equation}
  \sigma_{\ge2\mathrm{i}}
  \approx
  \frac12\sigma_{\mathrm{i}}^2
  \int \mathrm{d}^2b \, A(b)^2
  -\frac13\sigma_{\mathrm{i}}^3
  \int \mathrm{d}^2b \, A(b)^3
  +\mathcal{O}(\sigma_{\mathrm{i}}^4).
\end{equation}
By analogy, we have
\begin{equation}
  \sigma_{\ge1\mathrm{i},\ge1\mathrm{j}} \equiv \sum_{n,m=1} \sigma_{n\mathrm{i},m\mathrm{j}}\,,
\end{equation}
\begin{equation}
  \sigma_{\ge1\mathrm{i},\ge1\mathrm{j}}
  = \int
  \mathrm{d}^2b\left(1-\mathrm{e}^{-\sigma_{\mathrm{i}}A(b)}\right)
  \left(1-\mathrm{e}^{-\sigma_{\mathrm{j}}A(b)}\right),
\end{equation}
and
\begin{equation}
  \sigma_{\ge1\mathrm{i},\ge1\mathrm{j}}
  \approx
  \sigma_{\mathrm{i}}\sigma_{\mathrm{j}}
  \int \mathrm{d}^2b \, A(b)^2
  -\frac12\sigma_{\mathrm{i}}\sigma_{\mathrm{j}}(\sigma_{\mathrm{i}}+\sigma_{\mathrm{j}})
  \int \mathrm{d}^2b \, A(b)^3
  +\mathcal{O}(\sigma_{\mathrm{i,j}}^4).
\end{equation}

\subsubsection{Exclusive definition}

In the eikonal model the exclusive double-$\mathrm{i}$ and $\mathrm{ij}$
cross sections are given by
\begin{equation}
  \sigma_{2\mathrm{i}}
  = \int \mathrm{d}^2b \,
  \mbox{$\frac12$}(\sigma_{\mathrm{i}}A(b))^2 \,
  \mathrm{e}^{-\sigma_{\mathrm{i}}A(b)}
  \approx
  \frac12\sigma_{\mathrm{i}}^2
  \int \mathrm{d}^2b \, A(b)^2
  -\frac12\sigma_{\mathrm{i}}^3
  \int \mathrm{d}^2b \, A(b)^3
  +\mathcal{O}(\sigma_{\mathrm{i}}^4),
\end{equation}
and
\begin{equation}
  \sigma_{1\mathrm{i},1\mathrm{j}}
  = \int
  \mathrm{d}^2b\,\sigma_{\mathrm{i}}\sigma_{\mathrm{j}}\,A(b)^2\,
  \mathrm{e}^{-(\sigma_{\mathrm{i}}+\sigma_{\mathrm{j}})A(b)}
  \approx
  \sigma_{\mathrm{i}}\sigma_{\mathrm{j}}
  \int \mathrm{d}^2b \, A(b)^2
  -\sigma_{\mathrm{i}}\sigma_{\mathrm{j}}(\sigma_{\mathrm{i}}+\sigma_{\mathrm{j}})
  \int \mathrm{d}^2b \, A(b)^3
  +\mathcal{O}(\sigma_{\mathrm{i,j}}^4).
\end{equation}

\subsection{The effective cross section}

In all cases, we take Eq.~(\ref{eq:sigmaeffdef}) as the definition of
\seff.

\subsubsection{Inclusive definition}

\begin{equation}
  \seff =
  \frac{\sigma_{\textit{incl\/}\mathrm{i}}^2}{2\sigma_{\textit{incl\/}\mathrm{ii}}}
  =\frac{\sigma_{\textit{incl\/}\mathrm{i}}\,\sigma_{\textit{incl\/}\mathrm{j}}}
  {\sigma_{\textit{incl\/}\mathrm{ij}}}
  = \frac1{\int \mathrm{d}^2b \, A(b)^2}\,.
\end{equation}

\subsubsection{Alternative inclusive definition}

\begin{equation}
  \seff =
  \frac{\sigma_{\ge1\mathrm{i}}^2}{2\sigma_{\ge2\mathrm{i}}}
  = \frac{\left[\int \mathrm{d}^2b\left(1-\mathrm{e}^{-\sigma_{\mathrm{i}}A(b)}\right)\right]^2}
  {2\int \mathrm{d}^2b\left(1-\mathrm{e}^{-\sigma_{\mathrm{i}}A(b)}
  -\sigma_{\mathrm{i}}A(b)\mathrm{e}^{-\sigma_{\mathrm{i}}A(b)}\right)}\,.
\end{equation}
With this definition, \seff is process-dependent, since its value
depends on $\sigma_{\mathrm{i}}$. The two different definitions in
Eq.~(\ref{eq:sigmaeffdef}) give different results,
\begin{equation}
  \seff =
  \frac{\sigma_{\ge1\mathrm{i}}\sigma_{\ge1\mathrm{j}}}{\sigma_{\ge1\mathrm{i},\ge1\mathrm{j}}}
  = \frac{
    \left[\int \mathrm{d}^2b\left(1-\mathrm{e}^{-\sigma_{\mathrm{i}}A(b)}\right)\right]
    \left[\int \mathrm{d}^2b\left(1-\mathrm{e}^{-\sigma_{\mathrm{j}}A(b)}\right)\right]
    }{\int
  \mathrm{d}^2b\left(1-\mathrm{e}^{-\sigma_{\mathrm{i}}A(b)}\right)
  \left(1-\mathrm{e}^{-\sigma_{\mathrm{j}}A(b)}\right)}\,.
\end{equation}
Both cases become process independent in the limit of small cross
sections, but with different process-dependent corrections,
\begin{equation}
  \seff \approx
  \frac1{\int \mathrm{d}^2b \, A(b)^2}
  -\left(2-\frac23\,\frac{\int \mathrm{d}^2b \, A(b)^3}
  {\left[\int \mathrm{d}^2b \, A(b)^2\right]^2}\right)
  \sigma_{\mathrm{i}}
  +\mathcal{O}(\sigma_{\mathrm{i}}^2),
\end{equation}
\begin{equation}
  \seff \approx
  \frac1{\int \mathrm{d}^2b \, A(b)^2}
  -\left(1-\frac12\,\frac{\int \mathrm{d}^2b \, A(b)^3}
  {\left[\int \mathrm{d}^2b \, A(b)^2\right]^2}\right)
  (\sigma_{\mathrm{i}}+\sigma_{\mathrm{j}})
  +\mathcal{O}(\sigma_{\mathrm{i,j}}^2).
\end{equation}

\subsubsection{Exclusive definition}

\begin{equation}
  \seff =
  \frac{\sigma_{1\mathrm{i}}^2}{2\sigma_{2\mathrm{i}}}
  = \frac{\left[\int \mathrm{d}^2b \,
  A(b) \, \mathrm{e}^{-\sigma_{\mathrm{i}}A(b)}\right]^2}
  {\int \mathrm{d}^2b \,
  A(b)^2 \,
  \mathrm{e}^{-\sigma_{\mathrm{i}}A(b)}}\,.
\end{equation}
\begin{equation}
  \seff =
  \frac{\sigma_{1\mathrm{i}}\sigma_{1\mathrm{j}}}{\sigma_{1\mathrm{i},1\mathrm{j}}}
  = \frac{
    \left[\int \mathrm{d}^2b \, A(b) \, \mathrm{e}^{-\sigma_{\mathrm{i}}A(b)}\right]
    \left[\int \mathrm{d}^2b \, A(b) \, \mathrm{e}^{-\sigma_{\mathrm{j}}A(b)}\right]
    }{\int
  \mathrm{d}^2b\,A(b)^2\,
  \mathrm{e}^{-(\sigma_{\mathrm{i}}+\sigma_{\mathrm{j}})A(b)}}\,.
\end{equation}
\begin{equation}
  \seff \approx
  \frac1{\int \mathrm{d}^2b \, A(b)^2}
  -\left(2-\frac{\int \mathrm{d}^2b \, A(b)^3}
  {\left[\int \mathrm{d}^2b \, A(b)^2\right]^2}\right)
  \sigma_{\mathrm{i}}
  +\mathcal{O}(\sigma_{\mathrm{i}}^2),
\end{equation}
\begin{equation}
  \seff \approx
  \frac1{\int \mathrm{d}^2b \, A(b)^2}
  -\left(1-\frac{\int \mathrm{d}^2b \, A(b)^3}
  {\left[\int \mathrm{d}^2b \, A(b)^2\right]^2}\right)
  (\sigma_{\mathrm{i}}+\sigma_{\mathrm{j}})
  +\mathcal{O}(\sigma_{\mathrm{i,j}}^2).
\end{equation}
It is worth noting that the ratio of integrals of the matter
distribution appearing in these expressions is a dimensionless feature
of a given model that does not depend on the proton-radius-like
parameter of the model, varying between about 1.25 for a `black disc'
model to about 1.75 for an exponential model, so is always between 1
and~2. Therefore the sign of the correction between the standard
definition of \seff and the other definitions is different in different
cases.

\section{Comparison with Berezhnoy et al.}
\label{sect:Berezhnoy}

Another paper, Ref.~\cite{Berezhnoy:2012xq}, has also made comparisons
with the LHCb data of Ref.~\cite{Aaij:2012dz}. In their Table~I, they
show good agreement between their predictions and the LHCb data. Since
they use a value of $\seff=14.5\,\mathrm{mb}$ to make these predictions
this is surprising, since we have seen in Fig.~\ref{fig:seff} that the
LHCb data are consistent with a value of \seff a factor of~2 higher,
even after taking into account the factor of~2 coming from the sum over
charge conjugate modes.

In this Appendix, we consider the analysis of
Ref.~\cite{Berezhnoy:2012xq} and highlight the causes of their apparent
agreement with data.

\subsection{The single-inclusive cross section}

To set the notation, we begin with Eq.~(15) of \cite{Berezhnoy:2012xq}:
\begin{equation}
  \label{eq:15}
  \mbox{``}
  \sigma_i^{\textit{incl}} = \sigma_1p_i^{\mathrm{c}\vee\mathrm{\bar{c}}}
  +\sigma_2(2p_i^{\mathrm{c}\vee\mathrm{\bar{c}}}-(p_i^{\mathrm{c}\vee\mathrm{\bar{c}}})^2)
  \mbox{''}.
\end{equation}
Although they do not precisely define $\sigma_i^{\textit{incl}}$, from
this equation we can infer that it is what we call the alternative
inclusive definition~-- the cross section for one or more $i$s, and that
$\sigma_{1,2}$ are the exclusive cross sections for 1 and 2
$\mathrm{c\bar{c}}$ pairs respectively. Despite this,
\cite{Berezhnoy:2012xq} (Eq.~(20)) uses the inclusive formula,
\begin{equation}
  \mbox{``}
  \sigma_2 = \frac{\sigma_1^2}{2\seff} = 1.3\pm0.4\,\mathrm{mb}
  \mbox{''},
\end{equation}
and sets $\sigma_1$ equal to the theoretical prediction for the
inclusive charm cross section, $6.1\pm0.9\,\mathrm{mb}$ to obtain that
value of $\sigma_2$. It is evident that it is the values of
$\sigma_{1,2}$ that are used in the remainder of the analysis. In fact,
in a given model for the matter density, it is possible to obtain the
values of $\sigma_{\textit{incl\/}\mathrm{c}}$ and \seff from the values
of $\sigma_{1,2}$ and, in the form factor model used by
\cite{Seymour:2013qka} for example, these values correspond to
$\sigma_{\textit{incl\/}\mathrm{c}}\approx10\,\mbox{mb}$ and
$\seff\approx18\,\mathrm{mb}$. On the other hand, using the values that
\cite{Berezhnoy:2012xq} quotes,
$\sigma_{\textit{incl\/}\mathrm{c}}=6.1\,\mbox{mb}$ and
$\seff=14.5\,\mathrm{mb}$, in the same form factor model, they should
have used $\sigma_1=4.2\,\mathrm{mb}$ and $\sigma_2=0.72\,\mathrm{mb}$
in their calculation.

Like LHCb, \cite{Berezhnoy:2012xq} includes the sum over charge
conjugate modes throughout and therefore
$p_i^{\mathrm{c}\vee\mathrm{\bar{c}}}$ is the probability that a
$\mathrm{c\bar{c}}$ pair produces one or more $i$ or $\bar{i}$s. Thus,
in our notation, and assuming
$p^{\mathrm{c}}_{\mathrm{D}}=p^{\mathrm{\bar{c}}}_{\mathrm{\overline{D}}}$,
\begin{equation}
  p_i^{\mathrm{c}\vee\mathrm{\bar{c}}} =
  p^{\mathrm{c}}_{\mathrm{D}}+p^{\mathrm{\bar{c}}}_{\mathrm{\overline{D}}}
  -p^{\mathrm{c}}_{\mathrm{D}}p^{\mathrm{\bar{c}}}_{\mathrm{\overline{D}}}
  =2p^{\mathrm{c}}_{\mathrm{D}}-(p^{\mathrm{c}}_{\mathrm{D}})^2.
\end{equation}
It is evident that Eq.~(\ref{eq:15}) is a truncation at two scatters of
a sum that should extend over all numbers of scatters,
\begin{equation}
  \label{eq:alln}
  \sigma_i^{\textit{incl}} = \sum_n\sigma_n
  \Bigl(1-(1-p_i^{\mathrm{c}\vee\mathrm{\bar{c}}})^n\Bigr).
\end{equation}
Using the same form factor model again, and
$\sigma_{\textit{incl\/}\mathrm{c}}=6.1\,\mbox{mb}$ and
$\seff=14.5\,\mathrm{mb}$, one obtains
$\sigma_3=0.13\,\mathrm{mb}$. Although this gives a negligible
correction to the single-inclusive cross section, we will see below that
neglecting $\sigma_3$ from the double-inclusive cross section results in
a more significant error.

Finally, we can note that in practice the probabilities
$p_i^{\mathrm{c}\vee\mathrm{\bar{c}}}$ are small (on the percent level)
and hence it is actually a good approximation to neglect terms
suppressed by factors of $p_i^{\mathrm{c}\vee\mathrm{\bar{c}}}$. Hence
we can approximate Eq.~(\ref{eq:alln}) as
\begin{equation}
  \sigma_i^{\textit{incl}} \approx \sum_n\sigma_n
  \Bigl(np_i^{\mathrm{c}\vee\mathrm{\bar{c}}}\Bigr)
  =p_i^{\mathrm{c}\vee\mathrm{\bar{c}}}\sigma_{\textit{incl\/}\mathrm{c}}.
\end{equation}
We summarize the main points of our analysis of Eq.~(15) of
\cite{Berezhnoy:2012xq} as:
\begin{itemize}
  \item $\sigma_i^{\textit{incl}}$ is defined as the alternative
    inclusive cross section, but the difference between this and the
    conventional definition is small in practice.
  \item $\sigma_{1,2}$ are defined as the exclusive cross sections for 1
    and 2 partonic scatters to produce $\mathrm{c\bar{c}}$ pairs
    respectively, but the inclusive formula is used to calculate
    them. This results in a significant error in the final result.
  \item The sum over number of scatters is truncated at 2. This makes a
    small difference in practice.
\end{itemize}

\subsection{The double-inclusive cross section for same-sign pairs}

Eqs.~(17) and~(19) of \cite{Berezhnoy:2012xq} read:
\begin{eqnarray}
  \label{eq:17}
  \mbox{``}
  \sigma_{i,i}^{\textit{same}} &=&
  \sigma_2((p_{i,i}^{\mathrm{c}\wedge\mathrm{\bar{c}}})^2+2(p_{i,i}^{\mathrm{c}\wedge\mathrm{\bar{c}}})(p_i^{\mathrm{c}\vee\mathrm{\bar{c}}}-p_{i,i}^{\mathrm{c}\wedge\mathrm{\bar{c}}})+(p_i^{\mathrm{c}\vee\mathrm{\bar{c}}}-p_{i,i}^{\mathrm{c}\wedge\mathrm{\bar{c}}})^2/2)
  \mbox{''}, \\
  \label{eq:19}
  \mbox{``}
  \sigma_{i,j}^{\textit{same}} &=&
  \sigma_2(0.5(p_{i,j}^{\mathrm{c}\wedge\mathrm{\bar{c}}})^2+2p_{i,i}^{\mathrm{c}\wedge\mathrm{\bar{c}}}p_{j,j}^{\mathrm{c}\wedge\mathrm{\bar{c}}}
  +2p_{i,i}^{\mathrm{c}\wedge\mathrm{\bar{c}}}(p_j^{\mathrm{c}\vee\mathrm{\bar{c}}}-p_{i,j}^{\mathrm{c}\wedge\mathrm{\bar{c}}}-p_{j,j}^{\mathrm{c}\wedge\mathrm{\bar{c}}})+
  \nonumber\\&&
  +2p_{j,j}^{\mathrm{c}\wedge\mathrm{\bar{c}}}(p_i^{\mathrm{c}\vee\mathrm{\bar{c}}}-p_{i,j}^{\mathrm{c}\wedge\mathrm{\bar{c}}}-p_{i,i}^{\mathrm{c}\wedge\mathrm{\bar{c}}})+
  (p_i^{\mathrm{c}\vee\mathrm{\bar{c}}}-p_{i,j}^{\mathrm{c}\wedge\mathrm{\bar{c}}}-p_{i,i}^{\mathrm{c}\wedge\mathrm{\bar{c}}})
  (p_j^{\mathrm{c}\vee\mathrm{\bar{c}}}-p_{i,j}^{\mathrm{c}\wedge\mathrm{\bar{c}}}-p_{j,j}^{\mathrm{c}\wedge\mathrm{\bar{c}}}))
  \mbox{''}.
  \phantom{(B.9)}
\end{eqnarray}
Because they use the alternative inclusive definition (the cross
sections for two or more $i$s and one or more $i$ and one or more $j$),
their structure is complicated, but if we take the leading terms for
small probabilities, we see the structure more clearly:
\begin{eqnarray}
  \sigma_{i,i}^{\textit{same}} &\approx&
  \mbox{$\frac12$}(p_i^{\mathrm{c}\vee\mathrm{\bar{c}}})^2\sigma_2
  =2(p^{\mathrm{c}}_{\mathrm{D}})^2\sigma_2, \\
  \sigma_{i,j}^{\textit{same}} &\approx&
  p_i^{\mathrm{c}\vee\mathrm{\bar{c}}}p_j^{\mathrm{c}\vee\mathrm{\bar{c}}}\sigma_2
  =4p^{\mathrm{c}}_{\mathrm{D_1}}p^{\mathrm{c}}_{\mathrm{D_2}}\sigma_2.
\end{eqnarray}
It is evident that these expressions are truncations at two
scatters. They agree with our expectations, since they include sums over
charge conjugate modes, so the first is $\mathrm{DD}$ or
$\mathrm{\overline{D}\,\overline{D}}$, each of which can only come from two
$\mathrm{c\bar{c}}$ pairs in one way, while the second is
$\mathrm{D_1D_2}$ or $\mathrm{\overline{D}_1\overline{D}_2}$, each of
which can come from two $\mathrm{c\bar{c}}$ pairs in two possible ways.

However, for this double-inclusive cross section, the neglect of the
three-$\mathrm{c\bar{c}}$ cross section is more significant. One expects
\begin{eqnarray}
  \sigma_{i,i}^{\textit{same}} &\approx&
  2(p^{\mathrm{c}}_{\mathrm{D}})^2(\sigma_2+3\sigma_3+\ldots), \\
  \sigma_{i,j}^{\textit{same}} &\approx&
  4p^{\mathrm{c}}_{\mathrm{D_1}}p^{\mathrm{c}}_{\mathrm{D_2}}(\sigma_2+3\sigma_3+\ldots),
\end{eqnarray}
since there are three $\mathrm{c}$ pairs within a
three-$\mathrm{c\bar{c}}$ event. Since our estimate of $\sigma_3$ is
approximately six times smaller than $\sigma_2$, $3\sigma_3$ is an
$\sim50\%$ correction to $\sigma_2$.

We summarize the main points of our analysis of Eqs.~(17) and~(19) of
\cite{Berezhnoy:2012xq} as:
\begin{itemize}
  \item $\sigma_{i,i}^{\textit{same}}$ and
    $\sigma_{i,j}^{\textit{same}}$ are defined as alternative inclusive
    cross sections, but the difference between these and the
    conventionally defined ones are small in practice.
  \item $\sigma_2$ continues to be defined as the exclusive cross
    section for 2 partonic scatters, but the inclusive formula is used
    to calculate it. This results in a significant error in the final
    result.
  \item The sum over number of scatters is truncated at 2. This results
    in a significant error in the final result.
\end{itemize}

\subsection{The double-inclusive cross section for opposite-sign pairs}

Eqs.~(16) and~(18) of \cite{Berezhnoy:2012xq} read:
\begin{eqnarray}
  \label{eq:16}
  \mbox{``}
  \sigma_{i,i}^{\textit{diff}} &=&
  \sigma_1p_{i,i}^{\mathrm{c}\wedge\mathrm{\bar{c}}}+
  \sigma_2(2p_{i,i}^{\mathrm{c}\wedge\mathrm{\bar{c}}}-(p_{i,i}^{\mathrm{c}\wedge\mathrm{\bar{c}}})^2+(p_i^{\mathrm{c}\vee\mathrm{\bar{c}}}-p_{i,i}^{\mathrm{c}\wedge\mathrm{\bar{c}}})^2/2)
  \mbox{''}, \\
  \label{eq:18}
  \mbox{``}
  \sigma_{i,j}^{\textit{diff}} &=&
  \sigma_1p_{i,j}^{\mathrm{c}\wedge\mathrm{\bar{c}}}+
  \sigma_2(2p_{i,j}^{\mathrm{c}\wedge\mathrm{\bar{c}}}-(p_{i,j}^{\mathrm{c}\wedge\mathrm{\bar{c}}})^2
  +2p_{i,i}^{\mathrm{c}\wedge\mathrm{\bar{c}}}p_{j,j}^{\mathrm{c}\wedge\mathrm{\bar{c}}}
  +2p_{i,i}^{\mathrm{c}\wedge\mathrm{\bar{c}}}(p_j^{\mathrm{c}\vee\mathrm{\bar{c}}}-p_{i,j}^{\mathrm{c}\wedge\mathrm{\bar{c}}}-p_{j,j}^{\mathrm{c}\wedge\mathrm{\bar{c}}})+
  \nonumber\\&&
  +2p_{j,j}^{\mathrm{c}\wedge\mathrm{\bar{c}}}(p_i^{\mathrm{c}\vee\mathrm{\bar{c}}}-p_{i,j}^{\mathrm{c}\wedge\mathrm{\bar{c}}}-p_{i,i}^{\mathrm{c}\wedge\mathrm{\bar{c}}})+
  (p_i^{\mathrm{c}\vee\mathrm{\bar{c}}}-p_{i,j}^{\mathrm{c}\wedge\mathrm{\bar{c}}}-p_{i,i}^{\mathrm{c}\wedge\mathrm{\bar{c}}})
  (p_j^{\mathrm{c}\vee\mathrm{\bar{c}}}-p_{i,j}^{\mathrm{c}\wedge\mathrm{\bar{c}}}-p_{j,j}^{\mathrm{c}\wedge\mathrm{\bar{c}}}))
  \mbox{''}.
  \phantom{(B.9)}
\end{eqnarray}
Again, because they use the alternative inclusive definition, their
structure is complicated. To find the leading terms, we have to consider
the additional probabilities
$p_{i,i}^{\mathrm{c}\wedge\mathrm{\bar{c}}}$ and
$p_{i,j}^{\mathrm{c}\wedge\mathrm{\bar{c}}}$ appearing in these
equations. These are defined as the probabilities for a
$\mathrm{c\bar{c}}$ pair produced in a single partonic collision to
produce an $i$ and an $\bar{i}$, and an $i$ and a $\bar{j}$ or an
$\bar{i}$ and a $j$, respectively. Since these involve a correlation
between the $\mathrm{c}$ and the $\mathrm{\bar{c}}$, in our notation
they are
\begin{equation}
  p_{i,i}^{\mathrm{c}\wedge\mathrm{\bar{c}}} = \mathcal{C}(p^{\mathrm{c}}_{\mathrm{D}})^2,
  \qquad
  p_{i,j}^{\mathrm{c}\wedge\mathrm{\bar{c}}} = 2\mathcal{C}p^{\mathrm{c}}_{\mathrm{D_1}}p^{\mathrm{c}}_{\mathrm{D_2}}.
\end{equation}
On the other hand, Eq.~(21) of \cite{Berezhnoy:2012xq} states:
\begin{equation}
  \label{eq:21}
  \mbox{``}
  p_{i,i}^{\mathrm{c}\wedge\mathrm{\bar{c}}} \approx (p_i^{\mathrm{c}\vee\mathrm{\bar{c}}})^2
  \qquad
  p_{i,j}^{\mathrm{c}\wedge\mathrm{\bar{c}}} = 2p_i^{\mathrm{c}\vee\mathrm{\bar{c}}}p_j^{\mathrm{c}\vee\mathrm{\bar{c}}}
  \mbox{''}.
\end{equation}
This implies that they are taking the value of the correlation
coefficient, $\mathcal{C}$, to be equal to~4, without explicitly saying
so and without backing up this choice with a Monte Carlo study or
experimental measurement.

Taking the leading terms for small probabilities, we then obtain:
\begin{eqnarray}
  \sigma_{i,i}^{\textit{diff}} &\approx&
  (p_i^{\mathrm{c}\vee\mathrm{\bar{c}}})^2
  (\sigma_1+\mbox{$\frac52$}\sigma_2), \\
  \sigma_{i,j}^{\textit{diff}} &\approx&
  2p_i^{\mathrm{c}\vee\mathrm{\bar{c}}}p_j^{\mathrm{c}\vee\mathrm{\bar{c}}}
  (\sigma_1+\mbox{$\frac52$}\sigma_2),
\end{eqnarray}
which agrees with our expectation, provided $\mathcal{C}=4$ is assumed.
Although these expressions are again truncations at two scatters, since
they start already at one scatter, the neglect of three or more scatters
is a small correction.

We summarize the main points of our analysis of Eqs.~(16) and~(18) of
\cite{Berezhnoy:2012xq} as:
\begin{itemize}
  \item $\sigma_{i,i}^{\textit{diff}}$ and
    $\sigma_{i,j}^{\textit{diff}}$ are defined as alternative inclusive
    cross sections, but the difference between these and the
    conventionally defined ones are small in practice.
  \item $\sigma_{1,2}$ continue to be defined as the exclusive cross
    sections for 1 and 2 partonic scatters respectively, but the
    inclusive formula is used to calculate them. This results in a
    significant error in the final result.
  \item The sum over number of scatters is truncated at 2. This makes a
    small difference in practice.
  \item A correlation of $\mathcal{C}=4$ between the fragmentation
    products of the charm and anticharm in a single $\mathrm{c\bar{c}}$
    pair is assumed. This is crucial for fitting data, but is not
    justified in the paper.
\end{itemize}

\subsection{Summary}

The LHCb data on double open charm production are incompatible with
$\seff=14.5\,\mathrm{mb}$. The apparent agreement between the data and
the analysis of Ref.~\cite{Berezhnoy:2012xq}, which uses
$\seff=14.5\,\mathrm{mb}$, is a coincidence caused by:
\begin{itemize}
  \item Using the formula that defines \seff in terms of single- and
    double-inclusive charm cross sections to calculate the single- and
    double-exclusive charm cross sections. This results in a factor
    $\sim1.5$ error in $\sigma_1$ and $\sim1.8$ in $\sigma_2$.
  \item Truncating the formula for the double-inclusive cross section
    for same-sign pairs at two scatters. This results in a factor of
    $\sim1.5$ error in $\sigma_{i,i}^{\textit{same}}$ and
    $\sigma_{i,j}^{\textit{same}}$.
  \item Assuming without explicit justification a correlation of $\mathcal{C}=4$
    between the fragmentation products of the charm and anticharm in a
    single $\mathrm{c\bar{c}}$ pair.
\end{itemize}

\bibliographystyle{JHEP}	
\bibliography{doubleCpaper}

\providecommand{\href}[2]{#2}\begingroup\raggedright\begin{thebibliography}{10}

\bibitem{Abramowicz:2013iva}
H.~Abramowicz, P.~Bartalini, M.~B{\"a}hr, N.~Cartiglia, E.~Dobson, et~al., {\it
  {Summary of the Workshop on Multi-Parton Interactions (MPI@LHC 2012)}},
  \href{http://xxx.lanl.gov/abs/1306.5413}{{\tt arXiv:1306.5413}}.

\bibitem{Akesson:1986iv}
{\bf AFS} Collaboration, T.~{\AA}kesson et~al., {\it {Double parton scattering
  in p p collisions at $\sqrt{s} = 63$ GeV}},  {\em Z.Phys.} {\bf C34} (1987)
  163.

\bibitem{Alitti:1991rd}
{\bf UA2} Collaboration, J.~Alitti et~al., {\it {A Study of multi - jet events
  at the CERN anti-p p collider and a search for double parton scattering}},
  {\em Phys.Lett.} {\bf B268} (1991) 145--154.

\bibitem{Abe:1993rv}
{\bf CDF} Collaboration, F.~Abe et~al., {\it {Study of four jet events and
  evidence for double parton interactions in $p\bar{p}$ collisions at $\sqrt{s}
  = 1.8$ TeV}},  {\em Phys.Rev.} {\bf D47} (1993) 4857--4871.

\bibitem{Abe:1997xk}
{\bf CDF} Collaboration, F.~Abe et~al., {\it {Double parton scattering in
  $\bar{p}p$ collisions at $\sqrt{s} = 1.8$ TeV}},  {\em Phys.Rev.} {\bf D56}
  (1997) 3811--3832.

\bibitem{Abazov:2009gc}
{\bf D0} Collaboration, V.~M. Abazov et~al., {\it {Double parton interactions
  in photon+3 jet events in $p p$ bar collisions $\sqrt{s}=1.96$ TeV}},  {\em
  Phys.Rev.} {\bf D81} (2010) 052012,
  [\href{http://xxx.lanl.gov/abs/0912.5104}{{\tt arXiv:0912.5104}}].

\bibitem{Aad:2013bjm}
{\bf ATLAS} Collaboration, G.~Aad et~al., {\it {Measurement of hard
  double-parton interactions in $W\to\ell\nu+2$ jet events at $\sqrt{s}=7$ TeV
  with the ATLAS detector}},  {\em New J.Phys.} {\bf 15} (2013) 033038,
  [\href{http://xxx.lanl.gov/abs/1301.6872}{{\tt arXiv:1301.6872}}].

\bibitem{CMS:awa}
{\bf CMS} Collaboration, {\it {Study of observables sensitive to double parton
  scattering in W + 2 jets process in p-p collisions at $\sqrt{s} = 7$ TeV}}, .

\bibitem{Kulesza:1999zh}
A.~Kulesza and W.~J. Stirling, {\it {Like sign $W$ boson production at the LHC
  as a probe of double parton scattering}},  {\em Phys.Lett.} {\bf B475} (2000)
  168--175, [\href{http://xxx.lanl.gov/abs/hep-ph/9912232}{{\tt
  hep-ph/9912232}}].

\bibitem{Aaij:2011yc}
{\bf LHCb} Collaboration, R.~Aaij et~al., {\it {Observation of $J/\psi$ pair
  production in $pp$ collisions at $\sqrt{s}=7$ TeV}},  {\em Phys.Lett.} {\bf
  B707} (2012) 52--59, [\href{http://xxx.lanl.gov/abs/1109.0963}{{\tt
  arXiv:1109.0963}}].

\bibitem{Aaij:2012dz}
{\bf LHCb} Collaboration, R.~Aaij et~al., {\it {Observation of double charm
  production involving open charm in pp collisions at $\sqrt{s}=7$ TeV}},  {\em
  JHEP} {\bf 1206} (2012) 141, [\href{http://xxx.lanl.gov/abs/1205.0975}{{\tt
  arXiv:1205.0975}}].

\bibitem{Seymour:2013qka}
M.~H. Seymour and A.~Siodmok, {\it {Constraining MPI models using sigma
  effective and recent Tevatron and LHC Underlying Event data}},
  \href{http://xxx.lanl.gov/abs/1307.5015}{{\tt arXiv:1307.5015}}.

\bibitem{Berezhnoy:2012xq}
A.~V. Berezhnoy, A.~K. Likhoded, A.~V. Luchinsky, and A.~A. Novoselov, {\it
  {Double $c\bar{c}$ production at LHCb}},  {\em Phys.Rev.} {\bf D86} (2012)
  034017, [\href{http://xxx.lanl.gov/abs/1204.1058}{{\tt arXiv:1204.1058}}].

\bibitem{Treleani:2007gi}
D.~Treleani, {\it {Double parton scattering, diffraction and effective cross
  section}},  {\em Phys.Rev.} {\bf D76} (2007) 076006,
  [\href{http://xxx.lanl.gov/abs/0708.2603}{{\tt arXiv:0708.2603}}].

\bibitem{Bahr:2013gkj}
M.~B{\"a}hr, M.~Myska, M.~H. Seymour, and A.~Siodmok, {\it {Extracting sigma
  effective from the CDF gamma+3jets measurement}},  {\em JHEP} {\bf 1303}
  (2013) 129, [\href{http://xxx.lanl.gov/abs/1302.4325}{{\tt
  arXiv:1302.4325}}].

\bibitem{Vanya}
I.~Belyaev private communication.

\bibitem{Norrbin:1998bw}
E.~Norrbin and T.~Sjostrand, {\it {Production mechanisms of charm hadrons in
  the string model}},  {\em Phys.Lett.} {\bf B442} (1998) 407--416,
  [\href{http://xxx.lanl.gov/abs/hep-ph/9809266}{{\tt hep-ph/9809266}}].

\bibitem{Norrbin:2000zc}
E.~Norrbin and T.~Sjostrand, {\it {Production and hadronization of heavy
  quarks}},  {\em Eur.Phys.J.} {\bf C17} (2000) 137--161,
  [\href{http://xxx.lanl.gov/abs/hep-ph/0005110}{{\tt hep-ph/0005110}}].

\bibitem{LHCb:2012fb}
{\bf LHCb Collaboration} Collaboration, R.~Aaij et~al., {\it {Measurement of
  the D+/- production asymmetry in 7 TeV pp collisions}},  {\em Phys.Lett.}
  {\bf B718} (2013) 902--909, [\href{http://xxx.lanl.gov/abs/1210.4112}{{\tt
  arXiv:1210.4112}}].

\bibitem{Aaij:2012cy}
{\bf LHCb Collaboration} Collaboration, R.~Aaij et~al., {\it {Measurement of
  the Ds+ - Ds- production asymmetry in 7 TeV pp collisions}},  {\em
  Phys.Lett.} {\bf B713} (2012) 186--195,
  [\href{http://xxx.lanl.gov/abs/1205.0897}{{\tt arXiv:1205.0897}}].

\bibitem{Gaunt:2011rm}
J.~R. Gaunt, C.~H. Kom, A.~Kulesza, and W.~J. Stirling, {\it {Probing double
  parton scattering with leptonic final states at the LHC}},
  \href{http://xxx.lanl.gov/abs/1110.1174}{{\tt arXiv:1110.1174}}.

\end{thebibliography}\endgroup

\end{document}